\documentclass[aps,prd,floats,floatfix,amssymb,twocolumn,superscriptaddress]{revtex4-1}

\usepackage[top=2cm, bottom=2.5cm, outer=1.5cm, inner=1.5cm]{geometry}
\usepackage[T1]{fontenc}
\usepackage{physics}
\usepackage{braket}
\usepackage{amsmath}
\usepackage{mathtools}
\usepackage{amssymb}
\usepackage{amsfonts}
\usepackage{stackrel}
\usepackage{mathrsfs}
\usepackage{bigints}
\usepackage{bbm}
\usepackage{float}
\usepackage[usernames,dvipsnames,svgnames,table]{xcolor}

\usepackage{graphbox}
\usepackage{url}
\usepackage{hyperref}
\usepackage{graphicx,graphics}
\usepackage{ulem}
\usepackage{color}

\def\l@subsubsection#1#2{}

\begin{document}

\title{Thermal effects on a global monopole with Robin boundary conditions}

\author{Lissa de Souza Campos}
\email[]{lissa.desouzacampos01@universitadipavia.it}
\affiliation{Dipartimento di Fisica, Universit\`a degli Studi di Pavia, Via Bassi, 6, 27100 Pavia, Italy}
\affiliation{Istituto Nazionale di Fisica Nucleare, Sezione di Pavia, Via Bassi, 6, 27100 Pavia, Italy}

\author{Jo\~ao Paulo  M.  Pitelli}
\email[]{pitelli@unicamp.br}
\affiliation{Departamento de Matem\'atica Aplicada, Universidade Estadual de Campinas,
13083-859 Campinas, S\~ao Paulo, Brazil}%

\begin{abstract}

  Within quantum field theory on a global monopole spacetime, we study thermal effects on a naked singularity and its relation with boundary conditions. We first obtain the two-points functions for the ground state and for thermal states of a massive, arbitrarily-coupled, free scalar field compatible with Robin boundary conditions at the singularity. We then probe these states using a static Unruh-Dewitt particle detector. The transition rate is analyzed for the particular cases of massless minimally or conformally coupled fields at finite temperature. To interpret the detector's behavior, we compute the thermal contribution to the ground-state fluctuations and to the energy density. We verify that the behavior of the transition rate, the fluctuations and the energy density are closely intertwined. In addition, we find that these renormalized quantities remain finite at the singularity for, and only for, Dirichlet boundary condition.

\end{abstract}

\maketitle

\section{Introduction}

The singularity theorems from the 60's and 70's clarified that singularities are not just a consequence of highly symmetrical scenarios---they are endemic of General Relativity \cite{Senovilla:2006db}. Notwithstanding, it is not unreasonable to expect naked singularities to be absent in nature. On one hand, the cosmic censorship conjecture stipulates that singularities, in our universe, shall always be hidden by a horizon. On the other hand, naked singularities naturally emerge in some (theoretical) scenarios---white holes, cosmic strings, global monopoles \cite{Vilenkin:1986hg}. Utterly, the consensus is probably that the matter can only be fully resolved by a theory of quantum gravity. Be that as it may, semi-classical analysis such as quantum field theory on curved spacetimes can elucidate the path towards a better understanding of the interface between quantum physics and general relativity.

Semi-classical analyses have brought to light noteworthy phenomena. Amongst these, there are: Hawking radiation, which concerns the particle production constituting the final process in the life of a black hole \cite{Hawking:1974rv}; and the quantum dressing of a naked singularity due to backreaction effects. Regarding the latter, it has been shown that scalar perturbations of a negative mass BTZ black hole brings about a horizon of Planckian radius covering its, previously naked, singularity \cite{Casals:2016ioo}. Maybe nature indeed hinders naked singularities, but there might be more than one mechanism to do so besides the existence of a horizon. With Hawking radiation in mind, one may wonder if evaporation couldn't be another one.

Motivated by the above discussion, in this work we study thermal effects within quantum field theory on a naked singularity spacetime and their dependence on the admissible boundary conditions. Specifically, we consider a free, scalar quantum field theory on a global monopole spacetime. Global monopoles arise when a global symmetry is spontaneously broken~\cite{Vilenkin:1986hg}. According to grand unified theories, they can result from phase transitions in the early universe. A global monopole spacetime $\mathcal{M}$ is described by the line-element
\begin{align}
  \label{eq: metric Global monopole}
    ds^2 =-dt^2 + dr^2 + \alpha^2 r^2 d\theta^2 + \alpha^2 r^2 \sin^2\theta d\varphi^2,
\end{align}
where $t\in\mathbb{R}$ yields a global, timelike, irrotational Killing vector field $\partial_t$, $r\in(0,\infty)$, $\theta\in[0,\pi)$, and $\varphi\in[0,2\pi)$. The parameter $\alpha\in(0,1)$ gives rise to a solid angular deficit. In particular, the hypersurface $\theta=\frac{\pi}{2}$ corresponds to a cone with a deficit angle of $2\pi(1-\alpha)$. In these coordinates, the Ricci and the Kretschmann scalars are, respectively,
\begin{equation}
  \mathbf{R} = \frac{2(1-\alpha ^2)}{\alpha ^2 r^2}\quad \text{ and }\quad \mathbf{K} = \mathbf{R}^2
\end{equation}
and the metric associated to \eqref{eq: metric Global monopole} solves Einstein field equations  with a classical energy-momentum tensor whose only non-vanishing components are
\begin{equation}
    T_{tt} = -T_{rr} = \frac{\mathbf{R}}{2}.
\end{equation}
It follows that the singularity at $r\rightarrow0$ is naked, timelike and  of curvature type.  Accordingly, $\mathcal{M}$ is a static, geodesically-incomplete, non-globally hyperbolic spacetime on which the propagation of quantum fields depends on the choice of boundary condition at the singularity.

For the stablishment of a quantum field theoretical framework, the first step consists on obtaining physically-sensible two-point functions. Since global monopoles are not globally hyperbolic spacetimes, the Klein-Gordon equation gives rise to an initial-boundary value problem. That is, given suitable initial data on a spacelike surface, corresponding solutions, if they exist, are specified by boundary conditions, yielding inequivalent dynamics. Amongst the possible boundary conditions to be imposed, we restrict our attention to the ones that generate physically sensible dynamics in the sense of Ishibashi and Wald \cite{waldjmp, Ishibashi:2003jd}. Namely, we focus on the ones that bring about self-adjoint extensions of the radial part of the Klein-Gordon operator.

In Ref.~\cite{pitelli}, one of the authors showed that the sensible dynamics for a scalar field around a global monopole are prescribed by Robin boundary conditions at the singularity. Given this infinite class of non-equivalent dynamics, Ref.~\cite{Barroso:2018pjs} extended the analysis of Dirichlet quantum fields in the global monopole spacetime considered in \cite{Mazzitelli:1990zv} to include these non-trivial Robin boundary conditions. Following the same reasoning, we study thermal states for quantum fields with Robin boundary condition, generalizing the results of Ref.~\cite{Carvalho:2000hm},where only Dirichlet boundary condition was considered.

Our interest lies not only in investigating thermal effects on naked singularities, but particularly in studying them together with the impact of having different, inequivalent, physically-sensible dynamics engendered by the naked singularity. With the corresponding two-point functions in hands, we can probe these different quantum states within the particle detector approach. To this goal, we consider an Unruh-DeWitt detector following a static trajectory and interacting with a quantum state on a global monopole spacetime via a monopole-type Hamiltonian operator, see e.g. \cite{Louko:2007mu} and the references therein. In the infinite interaction time limit and up to first order perturbation theory, the instantaneous transition rate $\dot{\mathcal{F}}$ of the detector is the Fourier transform of the pull-back $\mathcal{G}(s)$ along the detector trajectory, parametrized by the proper time interval $s=\tau-\tau^\prime$, of the two-point function of the underlying field evaluated at the detector's energy gap $\Omega$:
 \begin{align}
  \label{eq: def transition rate}
  \dot{\mathcal{F}} &= \int_\mathbb{R}ds e^{-i\Omega s} \mathcal{G}(s).
 \end{align}
The transition rate characterizes the probabilities of excitations, for $\Omega>0$, and de-excitations, for $\Omega<0$, of the detector. In general, the transition rate can be seen as a function of the detector's trajectory, of its energy gap and, of course, of the quantum state to which it is coupled. We obtain expressions that can be easily studied numerically for an arbitrary parameter set, but we focus on the response of a detector coupled to thermal states of massless, minimally or conformally coupled fields. In addition, for a given state on a fixed global monopole background, we see $\dot{\mathcal{F}}$ merely as a function of the distance between the detector and the naked singularity by fixing an arbitrary energy gap $\Omega>0$, and we compare its behaviour for different deficit angles and different boundary conditions.

Last, to paint a better picture of the consequences of taking thermal effects together with different boundary conditions into account, we compute the thermal contributions to the expectation value of the field squared and the energy density of renormalized thermal states. Such quantities have been computed at the ground state with Dirichlet and Robin boundary conditions \cite{Mazzitelli:1990zv,Barroso:2018pjs}, and at thermal states with Dirichlet boundary condition \cite{Carvalho:2000hm}. Yet, as for the two-point functions, considering both thermality and general boundary conditions renders a novelty character to our work.

We proceed as follows. In Section \ref{sec: Two-Point functions}, we construct the two-point functions of the ground-state and of thermal states. Then, we obtain an analytic expression for the transition rate, in Section \ref{sec: Transition rate}, and we study it, numerically (by truncating an infinite sum) for the case of massless, minimally coupled fields at finite temperature. The analysis with conformal coupling is left to the Appendix. In Section \ref{sec: Discussion}, we summarize the results concerning the thermal fluctuations and the energy density of the renormalized thermal state and we discuss its relation with the behavior of the detector. Final remarks are included in Section \ref{sec: Conclusion}.

\section{Two-Point functions}
\label{sec: Two-Point functions}

In this section we obtain two-point functions of the ground state and of thermal states for a free, scalar, massive quantum field theory on a global monopole spacetime $\mathcal{M}$. We follow the same procedure as detailed in \cite{Dappiaggi:2016fwc,Dappiaggi:2018xvw,Bussola:2017wki,Campos:2020lpt}. First, in Section \ref{subsec: The Klein-Gordon equation}, we obtain the solutions of the Klein-Gordon equation by mode-expansion. Secondly, we invoke spectral theory of second-order partial differential operators to study the radial equation. Its Green function is unique up to the choice of a boundary condition at the naked singularity, as specified in Section \ref{subsec: Robin boundary conditions}. The symmetries of the spacetime together with the Klein-Gordon equation, the canonical commutation relations and the restriction to sensible dynamics completely determine the integral kernel of two-point functions of physically-sensible quasifree ground and thermal states of local Hadamard form. This is detailed in Section \ref{subsec: Ground and thermal states}, where we also write down their explicit expressions.

\subsection{The Klein-Gordon equation}
\label{subsec: The Klein-Gordon equation}

Let us consider a free scalar field $\Psi : \mathcal{M} \to \mathbb{R}$ with mass $m_0$, coupled to the scalar curvature by a coupling parameter $\xi\geq 0$. Its dynamics is described by the Klein-Gordon equation:
\begin{equation}
\label{eq: KG eq}
  P\Psi = (\Box - m_0^2 - \xi \mathbf{R}) \Psi = 0.
\end{equation}
We consider solutions that can be written in the form
\begin{align}
\label{eq: ansatz sol kg}
\Psi_{\omega,\ell}(t,r,\theta,\varphi) =  e^{-i\omega t } R(r)Y_\ell^m(\theta,\varphi),
\end{align}
where $Y_\ell^m(\theta,\varphi)$ are the spherical harmonics with eigenvalues $-\ell(\ell+1)$, while the function $R(r)$ satisfies the Bessel equation
\begin{equation}
  \label{eq: the radial equation}
    R''(r)+ \frac{2}{r}R'(r)+ \left(p^2 - \frac{\lambda_{\ell,\xi,\alpha}}{r^2} \right)R(r)=0,
\end{equation}
with
\begin{gather}
  p^2 := \omega^2 - m_0^2, \label{eq:def p(omega2)}\\
  \lambda_{\ell,\xi,\alpha} := \frac{\ell(\ell+1)+ 2\xi(1-\alpha^2)}{\alpha^2}.
\end{gather}

A basis of solutions $\{R_{1},R_{2}\}$ of \eqref{eq: the radial equation} is given in terms of the spherical Bessel functions of first and second kind, respectively $j_\nu$ and $y_\nu$:
\begin{align}
\label{eq: basis radialfunction}
&R_{1}(pr)= j_{\nu}(pr), \quad
R_{2}(pr) = p\, y_{\nu}(pr),
\end{align}
with index
\begin{align}
\label{eq: nu}
& \nu := \frac{-1 + \sqrt{1 + 4\lambda_{\ell,\xi,\alpha}}}{2}\geq0
\end{align}

The solutions \eqref{eq: basis radialfunction} are normalized in order to have a unit Wronskian, and their dependence on $p$ is made explicit for notational convenience. Clearly, any linear combination of \eqref{eq: basis radialfunction} solves the radial equation \eqref{eq: the radial equation}. However, not all of them yield self-adjoint extensions for the radial part of the Klein-Gordon operator. In the next section, we show how to restrict the space of radial solutions to a self-adjoint domain.

\subsection{Robin boundary conditions}
\label{subsec: Robin boundary conditions}

Self-adjoint extensions of a partial differential operator are in correspondence with its square-integrable solutions. Taking the radial equation as a Sturm-Liouville problem with eigenvalue $p^2$ singles-out the appropriate Hilbert space of solutions: the space of square-integrable functions with respect to the measure $q(r)=r^2$. By direct inspection, and taking into account Weyl's endpoint classification, we find that $r\rightarrow \infty$ is a limit-point, and that $r\rightarrow 0$ is limit-point for $\ell>0$, but limit-circle for $\ell=0$. By \cite[Thm.10.4.5]{Zettl:2005}, it follows that there is a one-parameter family of (generalized) Robin boundary conditions that can be chosen for $\ell=0$ at $r\rightarrow 0$ consistently with self-adjoint extensions.

Let us parametrize the Robin boundary conditions by $\gamma$ and let us define the auxiliary quantity $\gamma_{\ell}$ by
\begin{align}
\gamma_{\ell}:=
\begin{cases}
\gamma\in[0,\pi), \quad &\text{ if }\ell=0,\\
0, \quad &\text{ if }\ell>0.\\
\end{cases}
\end{align}
The most general solution that is square-integrable at the endpoint $r\rightarrow0$ and yields self-adjoint extensions for the radial part of the Klein-Gordon operator can be written as
\begin{align}
  \label{eq: R_gamma}
R_{\gamma_{\ell}}(pr):= \cos(\gamma_{\ell})R_{1}(pr) - \sin(\gamma_{\ell})R_{2}(pr).
\end{align}
Since $R_1$ is the principal solution, the self-adjoint extension determined by taking $\gamma=0$ corresponds to the Friedrichs extension and we refer to this particular case as Dirichlet boundary condition. Also, note that for $\ell>0$, the solution \eqref{eq: R_gamma} indeed reduces to the principal solution.

By standard methods of singular Sturm-Liouville theory \cite[Ch.10]{Zettl:2005}, we can construct the Green function of the radial equation \eqref{eq: the radial equation}. Following exactly the same procedure as in \cite{Dappiaggi:2016fwc,Dappiaggi:2018xvw,Bussola:2017wki,Campos:2020lpt} and invoking precisely the same symmetry arguments for performing the contour integration, the spectral resolution of the radial Green function gives rise to the following identity (mind that $p=p(\omega^2)$ as per Eq.~\eqref{eq:def p(omega2)})
\begin{align}
  \label{eq:this one to compare it}
\int\limits_{m_0^2}^\infty d\omega^2 \frac{p}{\pi}\frac{R_{\gamma_\ell}(pr) R_{\gamma_\ell}(pr')}{\cos(\gamma_\ell)^2+p^2 \sin(\gamma_\ell)^2} = - \frac{\delta(r-r')}{q(r)}.
\end{align}
The identity above fixes the integral kernel of the two-point functions, as we show in the next section.

\subsection{Ground and thermal states}
\label{subsec: Ground and thermal states}

A physically-sensible two-point function on a global monopole spacetime $\mathcal{M}$ is a positive bidistribution $\mathcal{G}_{\beta,\gamma} \in \mathcal{D}'(\mathcal{M}\times \mathcal{M})$ that solves the Klein-Gordon equation in each entry and is of local Hadamard form. Taking into account that $\mathcal{M}$ is static and spherically-symmetric, and given the addition formula for the spherical harmonics, we consider the following ansatz for the integral kernel of $\mathcal{G}_{\beta,\gamma}$:
 \begin{equation}\label{eq: G KMS}
  \mathcal{G}_{\beta,\gamma}(x,x') = \sum_{\ell=0}^\infty \int\limits_{0}^{\infty} d\omega \mathcal{T}_{\beta}(t,t')\mathcal{R}_{\gamma}(r,r')   \Xi(\theta,\varphi,\theta',\varphi'),
 \end{equation}
 \noindent where $x=(t,r,\theta,\varphi)\in\mathcal{M}$,
 \begin{align}
   \label{eq: angular part of G}
    \Xi(\theta,\varphi,\theta',\varphi'):=\frac{2\ell+1}{4\pi}& P_\ell ( \cos(\theta)\cos(\theta')+\nonumber\\&+\sin(\theta)\sin(\theta')\cos(\varphi-\varphi')),
 \end{align}
 and $P_\ell$ is the Legendre function of first kind.

 The time function $\mathcal{T}_{\beta}(t,t')$ specifies the support of $\mathcal{G}_{\beta,\gamma}$ with respect to the Fourier frequency $\omega$. In turn, its support specifies the nature of the corresponding state. The two-point function of a ground state has support over positive $\omega$ frequencies, hence we take
 \begin{equation}
   \label{eq: T(t,t') for ground state}
   \mathcal{T}_{\infty}(t,t')=e^{-i\omega(t-t'-i 0^+)}.
 \end{equation}
 The two-point function of a thermal state at inverse-temperature $\beta$ with respect to the Killing field $\partial_t$ is one that satisfies the KMS condition, see e.g. \cite{Bratteli:1996xq}. This property is guaranteed to hold when taking
\begin{equation}
  \label{eq: T(t,t') for KMS state}
  \mathcal{T}_{\beta}(t,t')= \frac{ e^{-i \omega (t-t'-i0^+)}}{1-e^{-\beta\omega}} + \frac{e^{+i \omega (t-t'+i0^+)} }{e^{\beta\omega}-1}.
\end{equation}

Note that $\mathcal{T}_{\infty}(t,t')$ is in fact the zero-temperature limit ($\beta\rightarrow\infty$) of $\mathcal{T}_{\beta}(t,t')$ and that the forms of both the time and the angular parts of ansatz \eqref{eq: G KMS} are restricted by the symmetries of the spacetime. However, the radial part $\mathcal{R}_{\gamma}(r,r')$ depends on the particular form of the metric. Specifically, it is related to the Green function of the radial part of the Klein-Gordon equation and it is uniquely determined, up to the choice of boundary conditions, by the canonical commutation relations, as we state in the following.

Analogously to the cases detailed in \cite{Dappiaggi:2016fwc,Dappiaggi:2018xvw,Bussola:2017wki,Campos:2020lpt}, it happens that ansatz \eqref{eq: G KMS} satisfies the canonical commutation relations provided the function $\mathcal{R}_{\gamma}$ is symmetric under the mapping $r\leftrightarrow r'$ and if
\begin{equation}\label{eq: cond for ccr2}
  \int\limits_{0}^\infty d\omega^2 \mathcal{R}_{\gamma}(r,r') = -\frac{\delta(r-r')}{\eta(r)},
\end{equation}
where $\eta(r)$ is such that $\eta(r)\sin(\theta) = \sqrt{|g|} = \alpha^2 r^2 \sin\theta$.
On the other hand, expression \eqref{eq: cond for ccr2} is closely related to the spectral resolution of the Green function of the radial equation. By comparing expressions \eqref{eq:this one to compare it} with \eqref{eq: cond for ccr2}, we directly obtain
\begin{equation}\label{eq: radial part of G}
  \mathcal{R}_{\gamma}(r,r') = \Theta(\omega-m_0)\frac{p}{\pi \alpha^2}\frac{R_{\gamma_\ell}(pr) R_{\gamma_\ell}(pr')}{\cos(\gamma_\ell)^2+p^2 \sin(\gamma_\ell)^2} .
\end{equation}

Two-point functions constructed as above are guaranteed to yield physically-sensible dynamics due to Wald and Ishibashi's work concerning static non-globally hyperbolic spacetimes \cite{Ishibashi:2003jd} and to be of local Hadamard form due to a general result by Sahlmann and Verch regarding the UV-behavior of ground and thermal states on static spacetimes \cite{sahlmann2000passivity}. Altogether, we conclude that two-point functions with integral kernel given by \eqref{eq: G KMS}, \eqref{eq: angular part of G} and \eqref{eq: radial part of G} characterize well-defined thermal states at inverse-temperature $\beta\in(0,\infty)$ with respect to the Killing field $\partial_t$ when $\mathcal{T}_{\beta}(t,t')$ is given by Eq.~\eqref{eq: T(t,t') for KMS state}. For $\mathcal{T}_{\beta}(t,t')$ given instead by the limiting case \eqref{eq: T(t,t') for ground state}, then \eqref{eq: G KMS} characterizes a ground-state and we shall denote it $\mathcal{G}_{\infty,\gamma}$.

We emphasize that even though expression \eqref{eq: G KMS} seems rather abstract, the integral can be analytically performed in some particular cases, e.g. for $\alpha\rightarrow 1$ and $\gamma=0$ it gives the standard closed-form expressions of Minkowski spacetime. Still, for general set of parameters, expression \eqref{eq: G KMS} is suitable for numerical analyses. In particular, when considering field fluctuations, numerical integration can be performed after taking the coincidence limit $x'\rightarrow x$ by invoking Lebesgue dominated convergence theorem.

\section{Transition rate}
\label{sec: Transition rate}

Consider an Unruh-DeWitt detector with energy gap $\Omega$ following a static trajectory of fixed spatial coordinates $(r, \theta,\varphi)$. The Fourier transform of $\mathcal{G}_{\infty,\gamma}$, given by \eqref{eq: G KMS} with \eqref{eq: T(t,t') for ground state}, along such trajectory gives the transition rate of the detector when coupled to the ground-state for an infinite proper time, as per \eqref{eq: def transition rate}:
\begin{widetext}
\begin{equation}
  \label{eq: transtition rate ground state}
\dot{\mathcal{F}}_{\infty,\gamma} (r)= \frac{\Theta(-\Omega-m_0)\sqrt{\Omega^2-m_0^2}}{2\pi \alpha^2}\sum_{\ell=0}^\infty    \frac{2\ell+1}{\cos(\gamma_\ell)^2+ (\Omega^2-m_0^2) \sin(\gamma_\ell)^2}   \left[R_{\gamma_\ell}\left(\sqrt{\Omega^2-m_0^2}r\right)\right]^2.
\end{equation}
\end{widetext}
Note that $\dot{\mathcal{F}}_{\infty,\gamma} (r)$ independs on the angular position of the detector and vanishes identically for excitations ($\Omega>0$), as expected to occur for a Boulware-like ground state. When coupled to a thermal state, as per \eqref{eq: G KMS} with \eqref{eq: T(t,t') for KMS state}, the transition rate reads instead
\begin{equation}
  \label{eq: transtition rate KMS state}
\dot{\mathcal{F}}_{\beta,\gamma } (r)= \frac{\text{sign}(\Omega)}{e^{\beta\Omega}-1}\left[\dot{\mathcal{F}}_{\infty} (r)\big|_{\Omega\mapsto -|\Omega|}\right].
\end{equation}

Both expressions \eqref{eq: transtition rate ground state} and \eqref{eq: transtition rate KMS state} hold for massive, arbitrarily-coupled fields and $\alpha\in(0,1]$. The $\ell$-sum can be analytically performed only in particular limiting cases. Yet, in any case, numerical analysis are easily performed due to the fast convergence of the sum in $\ell$. In this section we focus on the transition rate for a detector coupled to a thermal state of a massless, minimal coupled field, and we study its behaviour with respect to its distance from the naked singularity. For convenience, we leave the analysis of the conformal coupling case to the Appendix.

Before discussing the numerical analysis, let us present some analytical considerations. For now, set $\xi=0$. In the limit $\alpha\rightarrow 1$, the spacetime corresponds simply to Minkowski spacetime with a boundary at $r\rightarrow0$. Let us consider this scenario for the sake of comparison, denoting
\begin{equation}
  \label{eq: transtition rate KMS state MINK}
\dot{\mathcal{F}}_{\beta,\gamma}^{\text{Mink}} (r) := \lim\limits_{\alpha\rightarrow 1} \dot{\mathcal{F}}_{\beta,\gamma } (r).
\end{equation}
In addition, for $\gamma=0$ \eqref{eq: transtition rate KMS state MINK} yields the expected result on Minkowski spacetime with no boundary:
\begin{equation}
  \label{eq: transtition rate KMS state MINK dirichlet}
\dot{\mathcal{F}}_{\beta,0}^{\text{Mink}} (r) =  \frac{\Theta(|\Omega|-m_0) }{2\pi} \frac{\text{sign}(\Omega)\sqrt{\Omega^2-m_0^2}}{e^{\beta\Omega}-1} .
\end{equation}

Given that $\lim\limits_{r\rightarrow 0} j_{\nu}(pr) = \delta_{\nu,0}$ and that $\nu=0$ only if $\ell=0$ (and $\xi=0$), it follows that at $r\rightarrow 0$ only the $\ell=0$ mode contributes to $\dot{\mathcal{F}}_{\beta,\gamma}(\Omega)$. To put it in another way, in the $r\to 0$ limit the detector is only affected by s-waves. This is expected since we recover spherical symmetry as we approach the singularity. Consequently, for any $\gamma\geq0$, close to the naked singularity we have
\begin{equation}
  \label{eq: transtition rate KMS state MINK robin at zero}
\lim\limits_{r\rightarrow 0}  \dot{\mathcal{F}}_{\beta,\gamma}^{\text{Mink}} (r) =  \dot{\mathcal{F}}_{\beta,0}^{\text{Mink}} (r)\cdot c_{\gamma}
\end{equation}
where $ c_{\gamma}$ is a constant given by
\begin{equation}
  \label{eq: transtition rate KMS state MINK robin at zero constant}
c_{\gamma}:=\lim\limits_{r\rightarrow 0}   \frac{R_{\gamma_\ell}(p r)^2}{\cos(\gamma_\ell)^2+ (\Omega^2-m_0^2) \sin(\gamma_\ell)^2}\bigg|_{\ell=0}.
\end{equation}
For arbitrary $\alpha$ instead, it holds (see Eq.~\eqref{eq: transtition rate ground state}):
\begin{equation}
  \label{eq: limit r 0 transtition rate KMS state massless field}
\lim\limits_{r\rightarrow 0} \dot{\mathcal{F}}_{\beta,\gamma}(r)= \frac{1}{\alpha^2}\lim\limits_{r\rightarrow 0}  \dot{\mathcal{F}}_{\beta,\gamma}^{\text{Mink}} (r).
\end{equation}

The behavior of the transition rate as the detector approaches the naked singularity on a global monopole, as given by \eqref{eq: limit r 0 transtition rate KMS state massless field}, is analogous to that of a detector approaching a cosmic string \cite{Davies:1987th}.  The Unruh-deWitt detector around a cosmic string was studied in \cite{Davies:1987th} by considering Dirichlet boundary condition. There, the same behaviour for the transition rate was obtained, namely $$\lim\limits_{r\rightarrow 0} \dot{\mathcal{F}}_{\beta,0}(r)=\frac{1}{\alpha^2} \lim\limits_{r\rightarrow 0}  \dot{\mathcal{F}}_{\beta,0}^{\text{Mink}} (r).$$
Notice that Eq.~\eqref{eq: limit r 0 transtition rate KMS state massless field} gives a more general result since it does not depend on the choice of the boundary condition. We should emphasize that the admissible boundary conditions at the global monopole singularity $r\rightarrow0$ are much simpler than the ones at the cosmic string singularity $z=0$ (see~\cite{Kay}, for instance). This justifies our choice for the global monopole as a toy model for spacetimes with naked singularities.

In addition, for all $r$, the contribution from the $\ell=0$ mode is such that
\begin{equation}
  \label{eq: l equal 0 transtition comparison with mink}
 \dot{\mathcal{F}}_{\beta,\gamma}(r)\big|_{\ell=0} =  \frac{1}{\alpha^2} \dot{\mathcal{F}}_{\beta,\gamma}^{\text{Mink}} (r)\big|_{\ell=0} .
\end{equation}
\noindent Explicitly, for $m_0=0$, the transition \eqref{eq: transtition rate KMS state} simplifies to
\begin{equation}
  \label{eq: transtition rate KMS state massless field}
\dot{\mathcal{F}}_{\beta,\gamma}(r)= \frac{1}{2\pi \alpha^2}  \frac{\Omega}{e^{\beta\Omega}-1}   \sum_{\ell=0}^\infty \frac{(2\ell+1)\left[R_{\gamma_\ell}\left(|\Omega|r\right)\right]^2}{\cos(\gamma_\ell)^2+ \Omega^2 \sin(\gamma_\ell)^2}.
\end{equation}
Expression \eqref{eq: transtition rate KMS state massless field} with $\xi=0$ is the one we consider in the following numerical analysis. For that, we perform the sum in $\ell$ from zero up to $\ell_{\text{max}}$. It is worth mentioning that the complete analysis is available online, in a Mathematica notebook, at \cite{git_global_monopole}.

\subsection{With respect to $\alpha$}

\begin{figure}[H]
  \centering
   \hspace{-.33cm}\includegraphics[align=c,width=.5\textwidth]{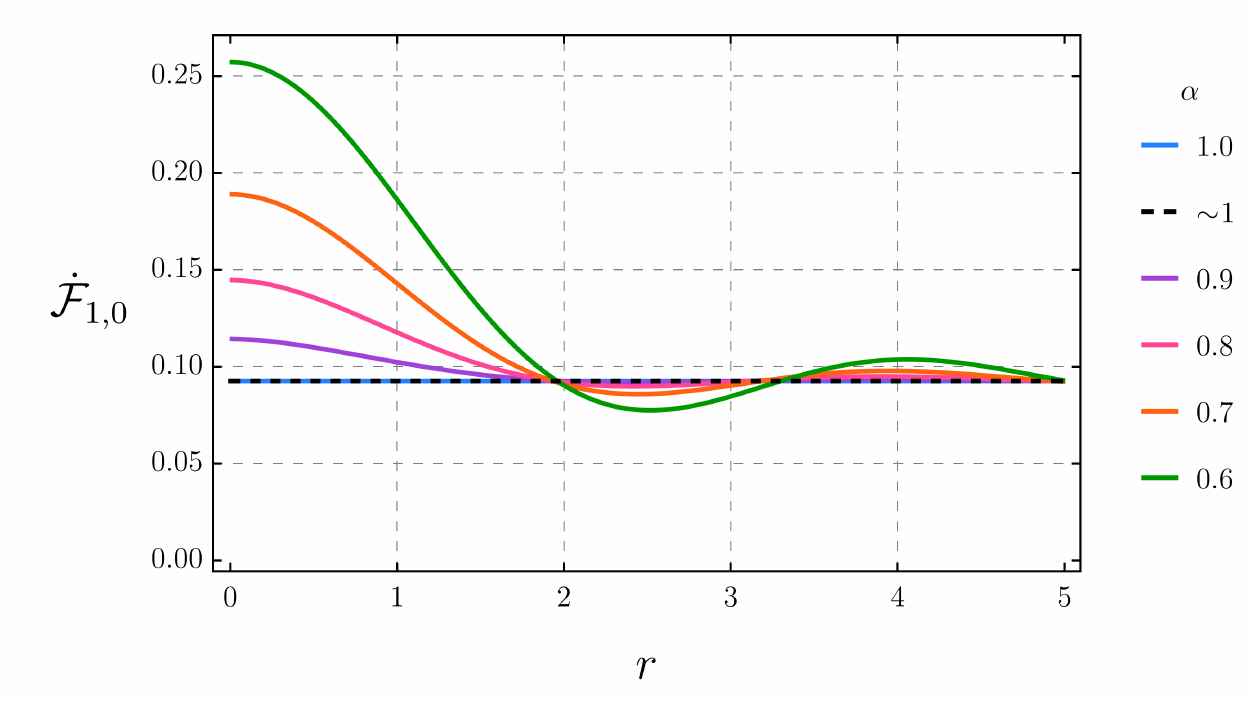}%
  \caption{The transition rate for the thermal state with $m_0=0$, $\xi=0$, $\beta=1$, $\Omega=1$, $\gamma=0$, $\ell_{\text{max}}=10$ and, from top to bottom with respect to the apex, $\alpha\in\{0.6,0.7,0.8,0.9,0.99999,1.0\}$. Note that the dashed line, for which $\alpha$ is close to $1$, shows essentially the same behavior as for $\alpha=1$. }
  \label{fig: transition rate KMS several alphas}
\end{figure}
Let us consider the Dirichlet case $\gamma=0$. As illustrated in Fig.~\ref{fig: transition rate KMS several alphas}, at large $r$, $\dot{\mathcal{F}}_{\beta,\gamma}(r)$ approximates, and oscillates around, the value of the transition rate in Minkowski, i.e.
\begin{equation}
  \label{eq: limit r infty transtition rate KMS state massless field}
\lim\limits_{r\rightarrow \infty} \dot{\mathcal{F}}_{\beta,\gamma}(r) \sim  \dot{\mathcal{F}}_{\beta,\gamma}^{\text{Mink}} (r) .
\end{equation}

\begin{figure}[H]
  \centering
  \hspace{-.33cm}\includegraphics[align=c,width=.5\textwidth]{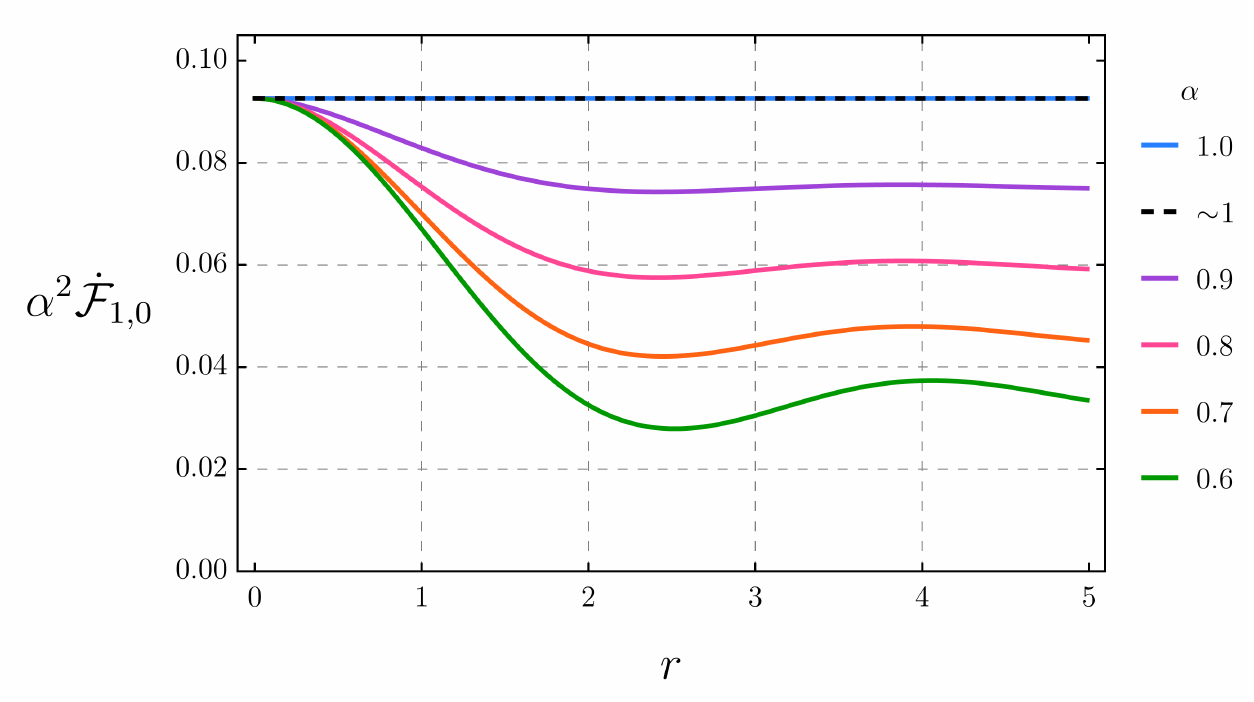}%
  \caption{The transition rate multiplied by $\alpha^2$ with the same parameters of Fig.~\ref{fig: transition rate KMS several alphas}.}
  \label{fig: transition rate times alpha2 ratio}
    \end{figure}

The behavior close to the singularity is most clear in Fig.~\ref{fig: transition rate times alpha2 ratio}, which is consistent with Eq.~\eqref{eq: limit r 0 transtition rate KMS state massless field}. Note that as $r\rightarrow 0$, all curves converge to the same value as that for $\alpha=1$.

\subsection{With respect to $\gamma$}
\begin{figure}[H]
              \centering
              \hspace{-.33cm}\includegraphics[align=c,width=.5\textwidth]{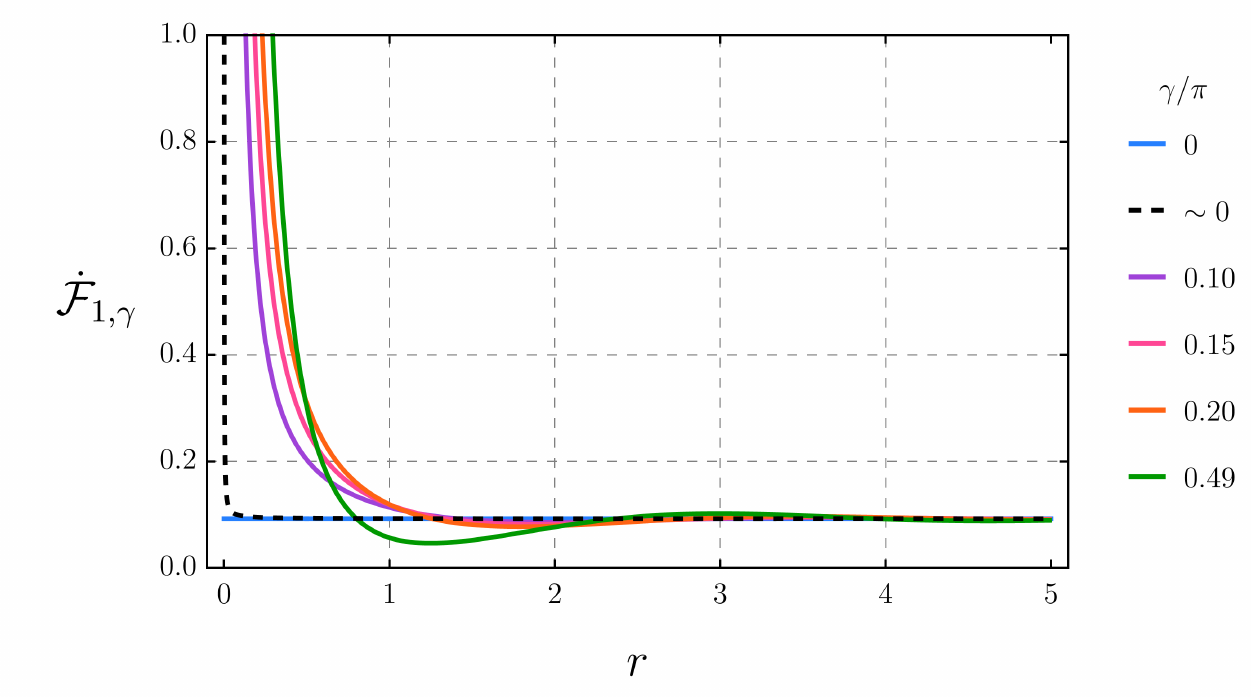}%
              \caption{The transition rate for the thermal state with $m_0=0$, $\xi=0$, $\beta=1$, $\Omega=1$, $\alpha=0.99999$, $\ell_{\text{max}}=10$ and several values of $\gamma$. The dashed line corresponds to $\gamma/\pi = 0.001$ and it shows a drastically different behavior as that of $\gamma=0$. }
              \label{fig: transition rate KMS several gammas}
                \end{figure}

Fig.~\ref{fig: transition rate KMS several gammas} encapsulates the main result we find. Namely, for any boundary condition $\gamma>0$, the transition rate diverges at $r\rightarrow 0$. In other words, only for Dirichlet boundary condition the spontaneous emission rate of a detector interacting with a thermal state remains finite at the naked singularity. In this case, we expect that the quantum fluctuations as well as the the energy supplied by the field must also be divergent. We show in the next section that this is indeed the case.

\section{Discussion}
\label{sec: Discussion}

Given that excitations cannot occur for a static detector coupled to a ground-state, the thermal contribution solely accounts for the spontaneous emission displayed by a detector coupled to a thermal state. To understand the behavior of the detector in the latter case, as summarized in the last section, we analyzed the thermal contribution to the ground-state fluctuations, denoted by $\Delta\mathcal{G}_{\beta,\gamma}(r)$, and the energy density, denoted by $E_{\beta,\gamma}(r)$, of the thermal state given by the time-time component of the renormalized energy-momentum tensor.

What we find is that the transition rate, the thermal fluctuations and the energy density are intertwined. For the minimally coupled case, as we approach the naked singularity, these three quantities diverge for $\gamma>0$, and are finite for $\gamma=0$. What is more, for $\ell=0$, the three quantities contrast with their counterparts in Minkowski spacetime in the same manner (as in Eq.~\eqref{eq: l equal 0 transtition comparison with mink}), viz.
\begin{equation}
  \label{eq: quantity vs quantity mink}
\{\text{quantity for general }\alpha\} = \frac{1}{\alpha^2} \{\text{quantity for }\alpha=1\}.
\end{equation}
Since the detector only sees the $\ell=0$ mode in the limit $r\rightarrow 0$, when $\xi=0$, relation \eqref{eq: quantity vs quantity mink} holds true when we consider such ``quantity'' to be the transition rate itself (summed up to $\infty$). In fact, that is also true for the thermal fluctuations. In the following, we summarize the analysis for $m_0=0$ and $\xi=0$.

\subsection{Thermal Fluctuations}
Let $\Delta\mathcal{G}_{\beta,\gamma}(x,x'):= \mathcal{G}_{\beta,\gamma}(x,x') -  \mathcal{G}_{\infty,\gamma}(x,x')$, where the two-point functions are given by \eqref{eq: G KMS} with \eqref{eq: T(t,t') for KMS state} and \eqref{eq: T(t,t') for ground state}, respectively. The thermal contribution to the ground-state fluctuations is given by
\begin{equation}
  \label{eq: def thermal fluctuations coincidence limit}
\Delta\mathcal{G}_{\beta,\gamma}(r):= \lim\limits_{x'\rightarrow x}  \left\{\Delta\mathcal{G}_{\beta,\gamma}(x,x')\right\}.
\end{equation}
Note that the coincidence limit above does not depend on the time and angular coordinates. For $m_0=0$ and $\xi=0$, we have
\begin{equation}
  \label{eq: thermal fluctuations coincidence limit explicit}
\Delta\mathcal{G}_{\beta,\gamma}(r) = \sum_{\ell=0}^\infty \int\limits_{0}^\infty d\omega  \frac{ (2\ell+1)}{2\pi^2 \alpha^2}   \frac{\omega}{e^{\beta\omega}-1} \frac{\left[R_{\gamma_\ell}\left(\omega r\right)\right]^2}{\cos(\gamma_\ell)^2+ \omega^2 \sin(\gamma_\ell)^2}.
\end{equation}
Defining $\Delta\mathcal{G}_{\beta,\gamma}^{\text{Mink}}$ as $\Delta\mathcal{G}_{\beta,\gamma}(r)$ with  $\alpha\rightarrow 1$, we find that
\begin{equation}
  \label{eq: limit r 0 thermal fluctuations comparison with mink}
 \lim\limits_{r\rightarrow 0} \Delta\mathcal{G}_{\beta,\gamma}(r)     =  \frac{1}{\alpha^2} \lim\limits_{r\rightarrow 0}\Delta\mathcal{G}_{\beta,\gamma}^{\text{Mink}} (r) ,
\end{equation}
and
\begin{equation}
  \label{eq: l equal 0 thermal fluctuations comparison with mink}
\Delta\mathcal{G}_{\beta,\gamma}(r)\big|_{\ell=0} =  \frac{1}{\alpha^2}\Delta\mathcal{G}_{\beta,\gamma}^{\text{Mink}} (r)\big|_{\ell=0} .
\end{equation}
That is, as we approach the singularity, the behavior of the thermal fluctuations, given by \eqref{eq: limit r 0 thermal fluctuations comparison with mink} is analogous to that of the transition rate, given by \eqref{eq: limit r 0 transtition rate KMS state massless field}. This behavior is illustrated by Figs. \ref{fig: transition rate times alpha2 ratio} and \ref{fig: thermal fluctuations times alpha2 KMS several gammas}, respectively.
The same holds for the $\ell=0$ mode, given expressions \eqref{eq: l equal 0 transtition comparison with mink} and \eqref{eq: l equal 0 thermal fluctuations comparison with mink}.
With respect to the boundary condition, the thermal fluctuations behave analogously to the transition rate, as in Fig.~\ref{fig: thermal fluctuations times alpha2 KMS several gammas}. That is, as $r\rightarrow 0$, $\Delta\mathcal{G}_{\beta,\gamma}(r)$ is finite if and only if $\gamma=0$.
\begin{figure}[H]
  \centering
  \hspace{-.33cm}\includegraphics[align=c,width=.5\textwidth]{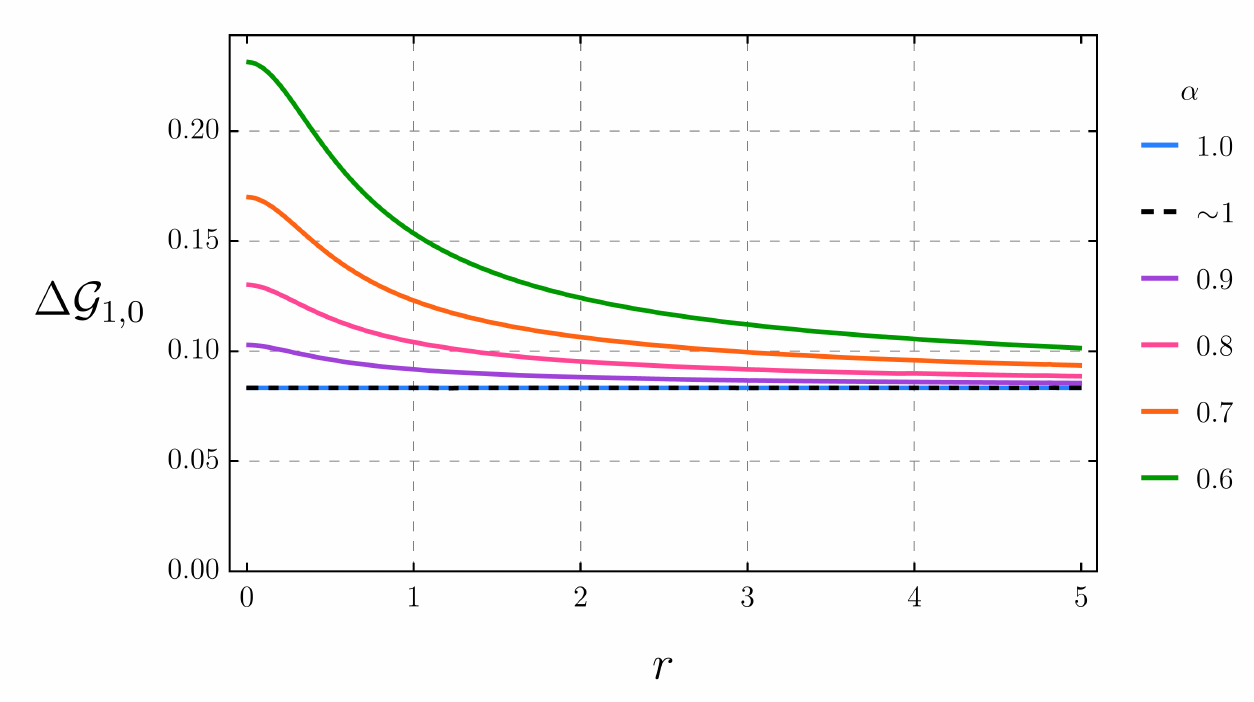}%
  \caption{Thermal fluctuations for $m_0=0$, $\xi=0$, $\beta=1$, $\gamma=0$, $\ell_{\text{max}}=50$ and, from top to bottom with respect to the apex, $\alpha\in\{0.6,0.7,0.8,0.9,0.99999,1.0\}$. Note that the behavior for $\alpha=0.99999$ (dashed line) and for $\alpha=1$ are indistinguishable.}
  \label{fig: thermal fluctuations KMS several alphas}
   \end{figure}
    \begin{figure}[H]
      \centering
      \hspace{-.33cm}\includegraphics[align=c,width=.5\textwidth]{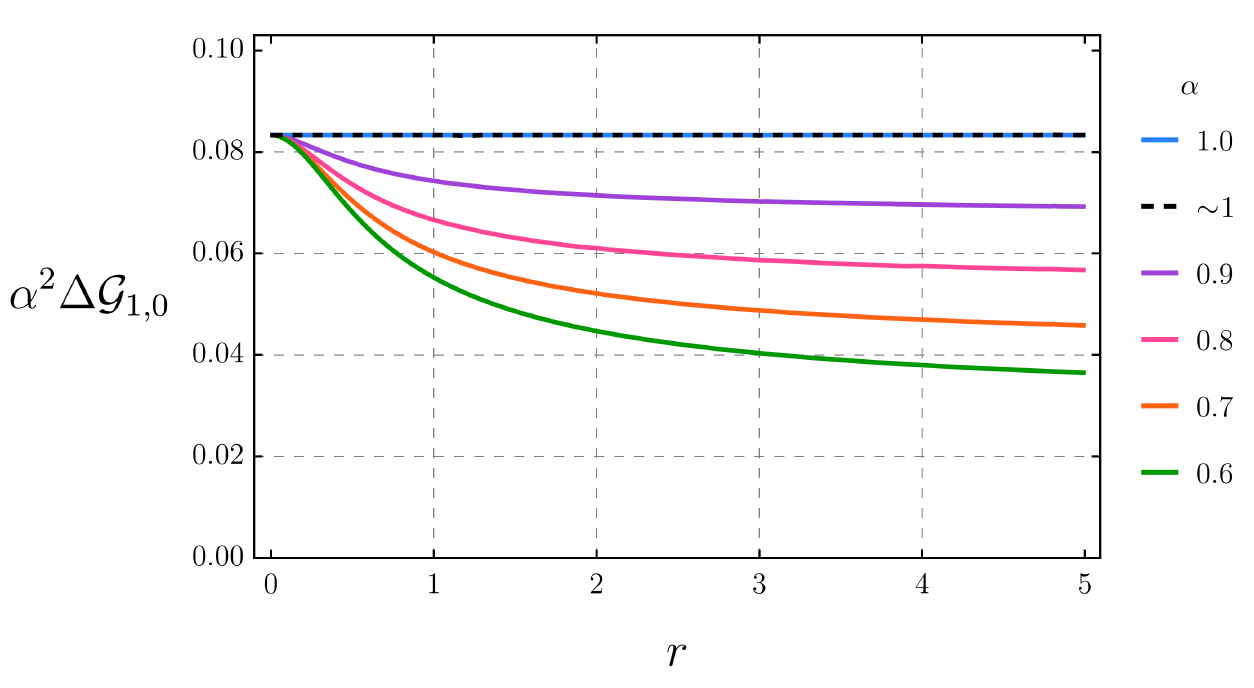}%
      \caption{Thermal fluctuations multiplied by $\alpha^2$ with the same parameters as in Fig.~\ref{fig: thermal fluctuations KMS several alphas}.}
      \label{fig: thermal fluctuations times alpha2 KMS several gammas}
        \end{figure}
\subsection{Energy density}
The energy-momentum tensor for the thermal state renormalized with respect to the ground-state is defined as, see \cite{Decanini:2005eg,Hack:2012qf} for details on Hadamard renormalization,
\begin{equation}
  \label{eq: renormalized stress energy tensor def}
    \braket{:T_{\mu\nu}(r):}_{\beta,\gamma} = \lim\limits_{x'\rightarrow x}\left\{\mathcal{D}_{\mu\nu}(x,x')\left[\Delta\mathcal{G}_{\beta,\gamma}(x,x')\right]\right\},
\end{equation}
where the differential operator $\mathcal{D}_{\mu\nu}(x,x')$ is given by
\begin{align}
  \label{eq: differential operator stress energy tensor moretti}
  &  \mathcal{D}_{\mu\nu}(x,x') := (1-2\xi) g_{\nu}\hspace{.5pt}^{\nu'}(x,x')\nabla_\mu\nabla_{\nu'} -2\xi\nabla_\mu\nabla_{\nu} + G_{\mu\nu} +\nonumber\\ & \quad + g_{\mu\nu}\left[  2\xi \Box + \left(2\xi-\frac{1}{2}  \right) g_{\rho}\hspace{.5pt}^{\rho'}(x,x') \nabla^\rho\nabla_{\rho'} -\frac{1}{2}m_0^2\right]  .
\end{align}
Accordingly, the energy density of the renormalized thermal state is simply the time-time component:
\begin{equation}
  \label{eq: energy density def}
  E_{\beta,\gamma}(r)   :=  \braket{:T_{00}(r):}_{\beta,\gamma}.
\end{equation}
For convenience, we omit the explicit expression for $E_{\beta,\gamma}(r)$, which is quite extensive and can be found in \cite{git_global_monopole}. In the following, we summarize the numerical results obtained for $E_{\beta,\gamma}(r)$ with $m_0=0$ and $\xi=0$.

\begin{figure}[H]
  \centering
  \hspace{-.33cm}\includegraphics[align=c,width=.5\textwidth]{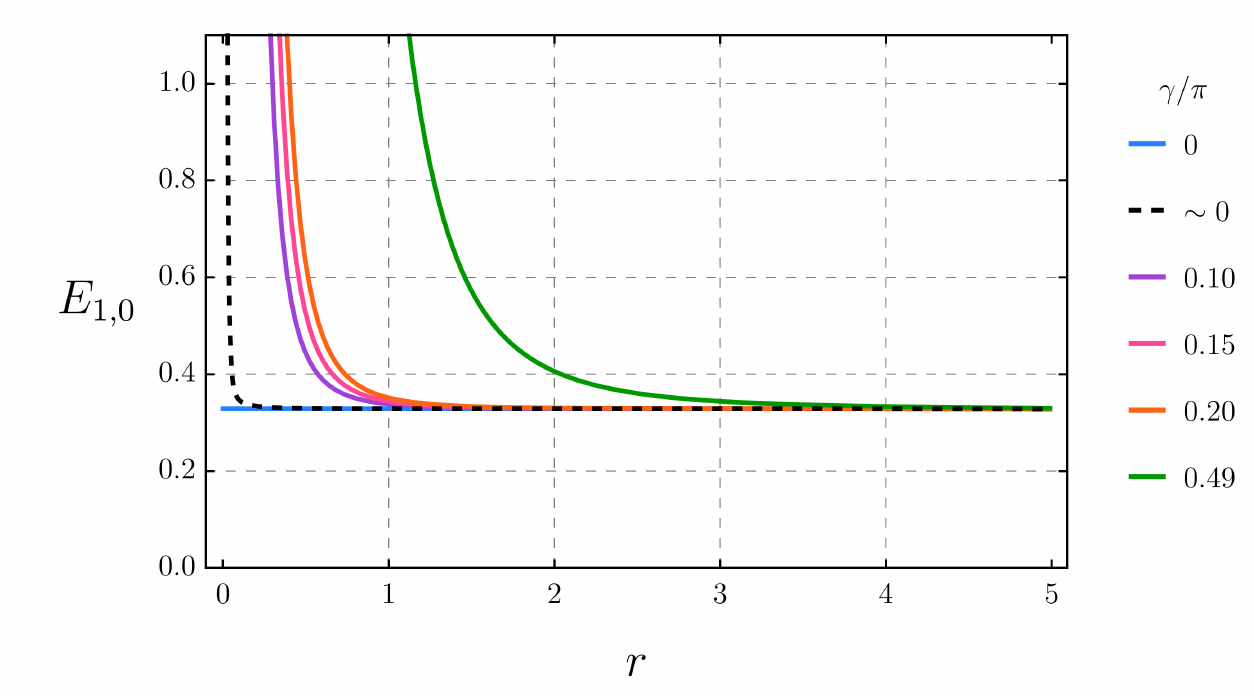}%
  \caption{Energy density for $m_0=0$, $\xi=0$, $\beta=1$,  $\alpha=0.99999$, $\ell_{\text{max}}=50$, and several values of $\gamma$. The dashed line corresponds to $\gamma/\pi = 0.001$.}
  \label{fig: energy density KMS several gammas}
\end{figure}

First and foremost, we find that the energy density converges at $r\rightarrow 0$ only for Dirichlet boundary condition, as shown in Fig.~\ref{fig: energy density KMS several gammas}. The divergence of the energy at the naked singularity is consistent with the divergence of spontaneous emission rate of the detector as in Fig.~\ref{fig: thermal fluctuations times alpha2 KMS several gammas}.

Moreover, we verified that, as it happens for the transition rate and for the thermal fluctuations, it holds that
\begin{equation}
  \label{eq: l equal 0 energy density comparison with mink}
      E_{\beta,\gamma}(r)\big|_{\ell=0} =  \frac{1}{\alpha^2}E_{\beta,\gamma}^{\text{Mink}}(r)\big|_{\ell=0}.
\end{equation}
\section{Conclusion}
\label{sec: Conclusion}

The propagation of quantum fields on non-globally hyperbolic spacetimes is not, in general, uniquely determined by the initial data on a spacelike surface. Wald and Ishibashi tackled this problem in Refs. \cite{waldjmp, Ishibashi:2003jd}, where they prescribed a way of extracting sensible dynamics for such fields by finding the positive self-adjoint extensions of the spatial component of the differential wave operator. These self-adjoint extensions, in turn, are prescribed by appropriate boundary conditions at the boundaries of the spacetime. Any quantity extracted from the quantum fields depends crucially on the choice of the boundary condition.

In this paper we investigated how the boundary condition at the classical singularity $r\rightarrow0$ of the global monopole spacetime affects the transition rate as measured by a particle detector. These (Robin) boundary conditions turns out to be extremely simple to handle, which makes the global monopole spacetime a very attractive toy model in the study of quantum effects due to naked singularities.

We considered a static Unruh-deWitt detector at a distance $r$ from the singularity. For massless and minimally coupled scalar fields on its ground state, the rate of excitation is zero in the infinite time interaction limit. This is expected for inertial observers on a general static spacetime. However, when the global monopole is immersed on a thermal bath with temperature $T\sim 1/\beta$, the induced extra thermal fluctuations creates a non-trivial scenario for the excitation of the detector. These thermal fluctuations are finite at the singularity $r\to 0$ only for Dirichlet boundary condition $\gamma=0$ and diverges for any other Robin boundary condition given $\gamma>0$. The transition rate behaves similarly in this limit. Nevertheless, the expected Minkowski thermal fluctuations are recovered in the $r\to \infty$ limit regardless of the choice of the boundary condition. This, in turn, induces the usual Minkowski transition rate for the quantum field on a thermal state. Moreover, for any boundary condition parametrized by $\gamma\in[0,\pi)$, we have
\begin{equation}
\frac{\dot{\mathcal{F}}_{\beta,\gamma}(0^+)}{\dot{\mathcal{F}}_{\beta,\gamma}^{\text{Mink}} (0^+)}=  \frac{\Delta\mathcal{G}_{\beta,\gamma}(0^+)}{\Delta\mathcal{G}_{\beta,\gamma}^{\text{Mink}} (0^+)}=\frac{E_{\beta,\gamma}(0^+) }{E_{\beta,\gamma}^{\text{Mink}} (0^+)}=\frac{1}{\alpha^2},
\end{equation}
\noindent which shows that the transition rate, the thermal fluctuations and the energy density of the renormalized thermal state are amplified by the presence of the singularity in exactly the same way.

The situation is rather different for conformally coupled fields. Relation \eqref{eq: quantity vs quantity mink} do not hold in general. Most importantly, the energy density diverges at $r\rightarrow 0$ for all $\gamma\geq 0$, and yet both the thermal fluctuations and the transition rate vanish. That is, even if there is an infinite amount of energy available, the detector will not undergo an excitation if the quantum field is not fluctuating. In the Appendix, we include the numerical analysis for the conformal coupling case.

Finally, in this work we showed that naked singularities manifest thermal effects, with a non-trivial behavior with respect to the admissible boundary conditions, in a static scenario. In a future work, it would be most interesting to study a dynamical scenario, either considering an accelerated detector or a model of collapse, in order to approach the question of whether naked singularities can evaporate with a bit more accurately.

\section*{Acknowledgments}

The work of L.S.C, who is grateful for discussions with C. Dappiaggi, was supported by a PhD scholarship of the University of Pavia.
\newpage
\appendix*
\section{Conformal Coupling}
\label{sec: Conformal Coupling}

\begin{figure}[H]
  \centering
   \hspace{-.33cm}\includegraphics[align=c,width=.5\textwidth]{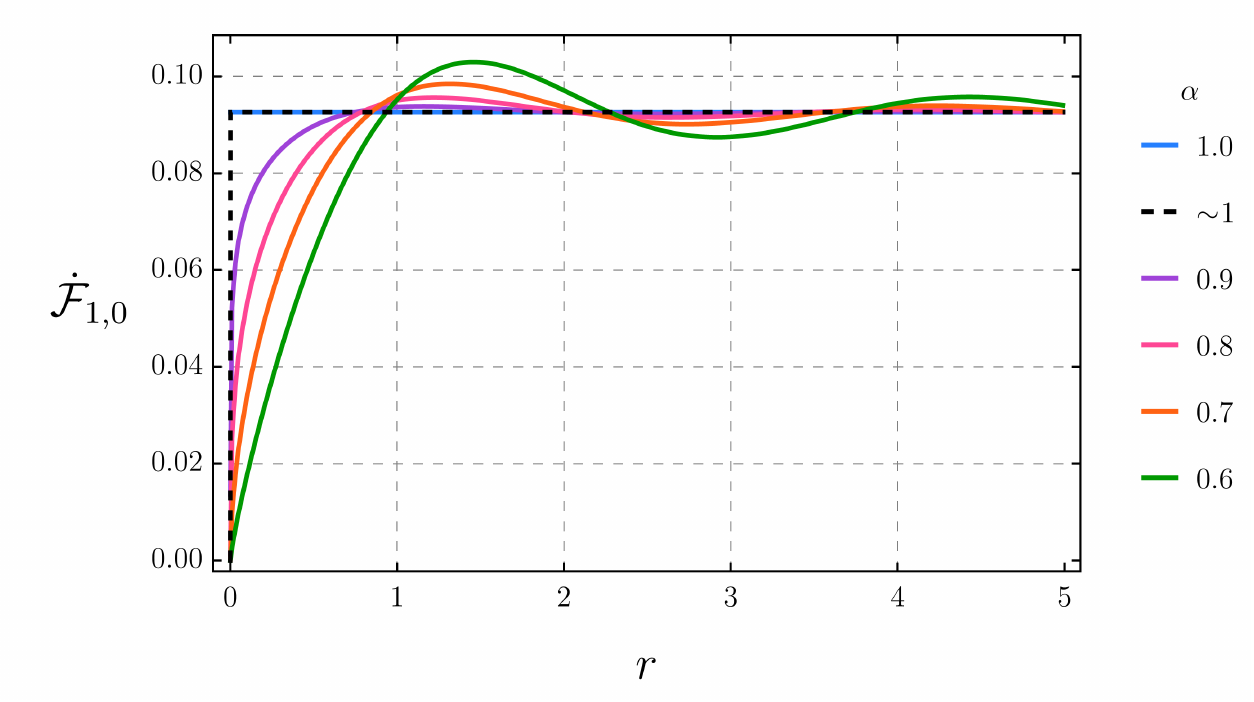}%
  \caption{The transition rate for the thermal state with $m_0=0$, $\xi=\frac{1}{6}$, $\beta=1$, $\Omega=1$, $\gamma=0$, $\ell_{\text{max}}=10$ and several $\alpha$'s. }
  \label{fig: transition rate KMS several alphas conformal}
\end{figure}

    For $\xi\neq 0$, $\alpha\in(0,1)$, and $\gamma\geq 0$, the parameter $\nu$ is never zero. This implies that in the limit $r\rightarrow 0$, even the $\ell=0$ contribution vanishes and it holds
    \begin{equation}
      \label{eq: transtition rate KMS state MINK robin at zero CONFORMAL}
    \lim\limits_{r\rightarrow 0}  \dot{\mathcal{F}}_{\beta,\gamma}(r) = 0.
    \end{equation}
    It follows that, as a function of $r$ for several $\alpha$'s, the behavior of the transition rate illustrated in Fig.~\ref{fig: transition rate KMS several alphas conformal} is quite different from the minimally coupled case as in Fig.~\ref{fig: transition rate KMS several alphas}. However, for large $r$, $\dot{\mathcal{F}}_{\beta,\gamma}(r)$ also approximates, oscillating around, its respective value on Minkowski spacetime. The plot for several boundary conditions is analogous to Fig.~\ref{fig: transition rate KMS several gammas} in the sense that for $\gamma>0$ the transition rate diverges at the singularity, with the only difference being that in this case the limit $r\rightarrow 0$ for $\gamma$ vanishes.

\begin{figure}[H]
  \centering
   \hspace{-.33cm}\includegraphics[align=c,width=.5\textwidth]{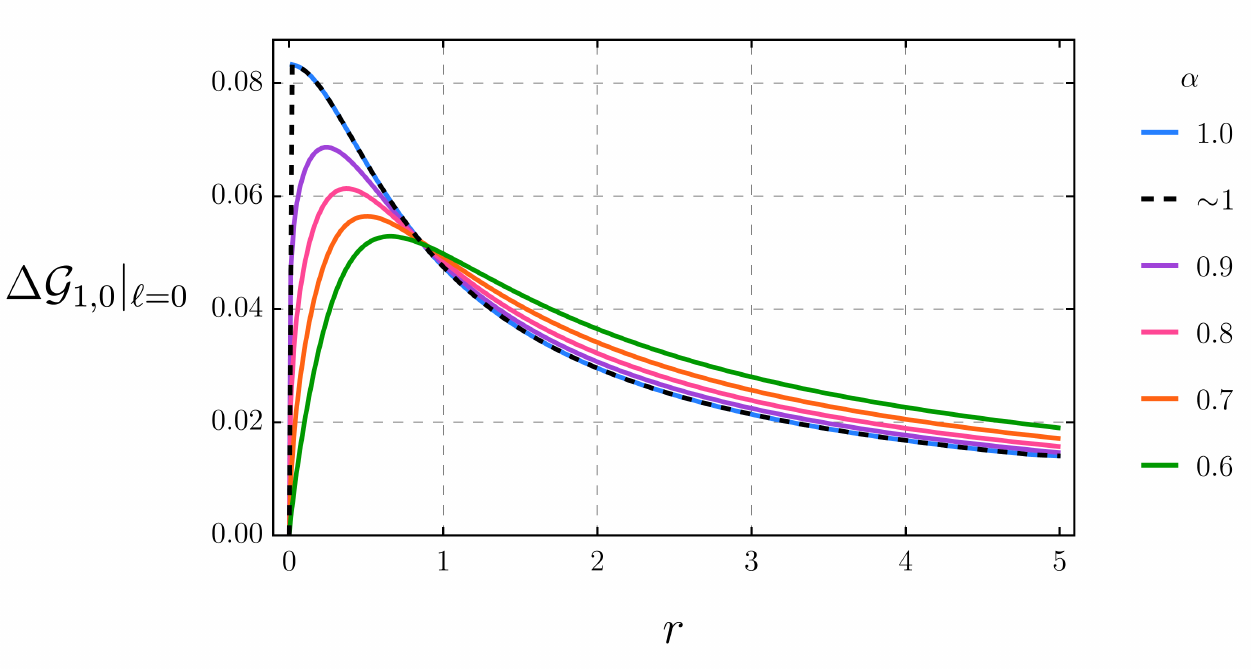}%
  \caption{Thermal Fluctuations for $m_0=0$, $\xi=\frac{1}{6}$, $\beta=1$, $\gamma=0$, $\ell_{\text{max}}=0$ and several $\alpha$'s.}
  \label{fig: app conformal thermal fluctuations l0}
\end{figure}

\begin{figure}[H]
  \centering
   \hspace{-.33cm}\includegraphics[align=c,width=.5\textwidth]{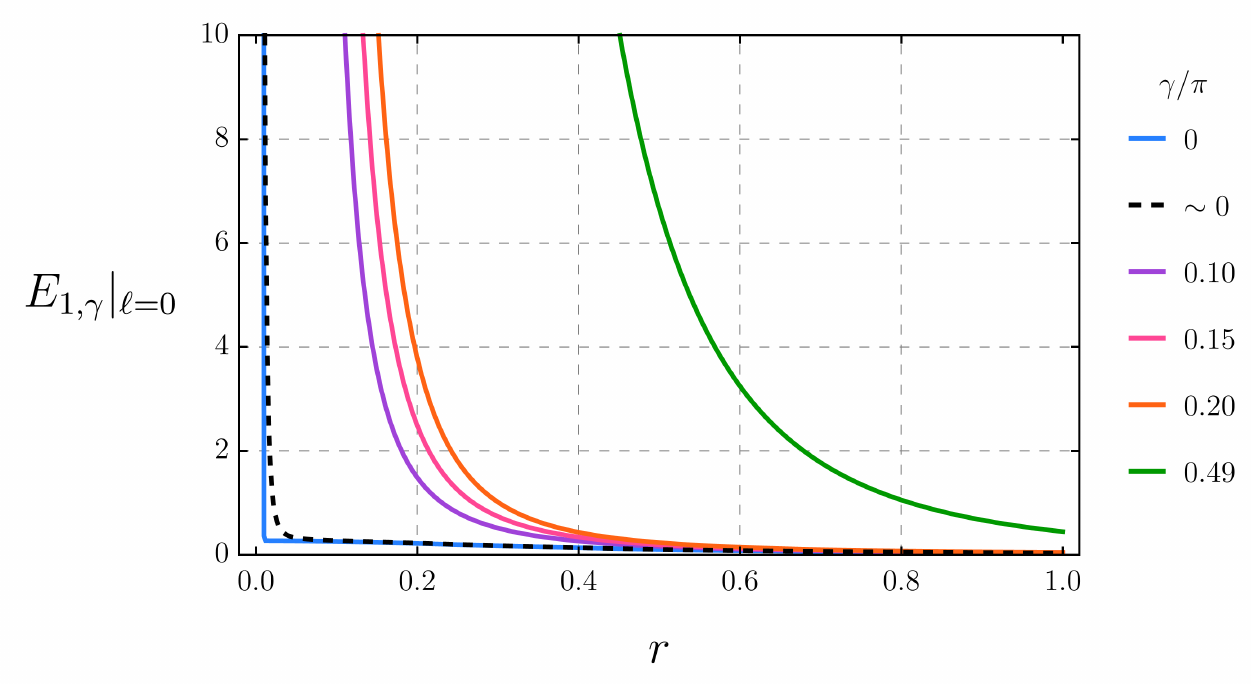}%
  \caption{Energy density for $m_0=0$, $\xi=\frac{1}{6}$, $\beta=1$, $\alpha=0.99999$, $\ell_{\text{max}}=0$ and several $\gamma$'s.}
  \label{fig: app conformal energy l0}
\end{figure}

For the same reason that the transition rate vanishes, so do the thermal fluctuations \eqref{eq: def thermal fluctuations coincidence limit} vanish in the limit $r\rightarrow 0$. What is most interesting in the conformally-coupled scenario is that the energy density actually diverges at the naked singularity---even for Dirichlet boundary condition---and yet, the transition rate vanishes there. For convenience, we illustrate the behavior 
of the thermal fluctuations, in Fig.~\ref{fig: app conformal thermal fluctuations l0}, and of the energy density, in Fig.~\ref{fig: app conformal energy l0}, considering only the $\ell=0$ mode.

\end{document}